\documentclass[a4paper,fleqn]{cas-dc}

\usepackage[numbers]{natbib}
\usepackage{newfloat}
\usepackage{listings}
\usepackage{amssymb}
\usepackage{makecell}
\usepackage{multirow}
\usepackage{booktabs}
\usepackage{graphicx}
\usepackage{subfigure}
\usepackage{algorithm}
\usepackage{algorithmic}
\usepackage{amsmath}
\usepackage{array}
\usepackage{url}
\usepackage{graphicx}
\usepackage{makecell}
\usepackage{multirow}
\usepackage{enumitem}
\usepackage{pifont}
\usepackage{glossaries}
\def\tsc#1{\csdef{#1}{\textsc{\lowercase{#1}}\xspace}}
\tsc{WGM}
\tsc{QE}
\begin{document}
\let\WriteBookmarks\relax
\def\floatpagepagefraction{1}
\def\textpagefraction{.001}

\shorttitle{Privacy Computing Meets Metaverse: Necessity, Taxonomy and
Challenges}    

\shortauthors{Chuan Chen, Yuecheng Li et al.}  

\title [mode = title]{Privacy Computing Meets Metaverse: Necessity, Taxonomy and Challenges}  

% % Title footnote mark
% % eg: \tnotemark[1]
% \tnotemark[<tnote number>] 

% % Title footnote 1.
% % eg: \tnotetext[1]{Title footnote text}


\affiliation[1]{organization={School of Computer Science and Engineering},
            addressline={Sun Yat-sen University}, 
            city={GuangZhou},
%          citysep={}, % Uncomment if no comma needed between city and postcode
            %postcode={000000}, 
            %state={},
            country={China}
            }

\affiliation[2]{organization={School of Software Engineering},
            addressline={Sun Yat-sen University}, 
            city={ZhuHai},
%          citysep={}, % Uncomment if no comma needed between city and postcode
            %postcode={000000}, 
            %state={},
            country={China}
            }

\affiliation[3]{organization={School of Automation},
            addressline={Guangdong University of Technology}, 
            city={GuangZhou},
%          citysep={}, % Uncomment if no comma needed between city and postcode
            %postcode={000000}, 
            %state={},
            country={China}
            }

%\author[<aff no>]{<author name>}[<options>]
\author[1]{Chuan Chen}
%\cormark[<corr mark no>]

\fnmark[1]

\ead{chenchuan@mail.sysu.edu.cn}

% \ead[url]{<URL>}

\credit{Conceptualization; Writing – review \& editing; Supervision; Resources}

%\author[<aff no>]{<author name>}[<options>]
\author[1]{Yuecheng Li}
% Footnote of the second author
\fnmark[2]
% Email id of the second author
\ead{liych78@mail2.sysu.edu.cn}
\credit{Investigation; Project administration; Writing – original draft; Visualization}

\author[1]{Zhenpeng Wu}
\fnmark[3]
\ead{iswuzp@163.com}
\credit{Investigation; Writing – original draft; Visualization}

\author[1]{Chengyuan Mai}
\fnmark[4]
\ead{maichy7@mail2.sysu.edu.cn}
\credit{Investigation; Project administration; Writing – original draft; Visualization}

\author[1]{Youming Liu}
\fnmark[5]
\ead{liuym66@mail2.sysu.edu.cn}
\credit{Investigation; Writing – original draft}

\author[1]{Yanming Hu}
\fnmark[6]
\ead{huym27@mail2.sysu.edu.cn}
\credit{Investigation; Writing – original draft}

\author[2]{Zibin Zheng**}
\fnmark[7]
\ead{zhzibin@mail.sysu.edu.cn}
\credit{Supervision}

\author[3]{Jiawen Kang}
\fnmark[8]
\ead{kavinkang@gdut.edu.cn}
\credit{Writing – review \& editing}

\cortext[2]{Corresponding author}

%\fntext[1]{xiaoming xiangde}


\begin{abstract}
	Metaverse, the core of the next-generation Internet, is a computer-generated holographic digital environment that simultaneously combines spatio-temporal, immersive, real-time, sustainable, interoperable, and data-sensitive characteristics. It cleverly blends the virtual and real worlds, allowing users to create, communicate, and transact in virtual form. With the rapid development of emerging technologies including augmented reality, virtual reality and blockchain, the metaverse system is becoming more and more sophisticated and widely used in various fields such as social, tourism, industry and economy. However, the high level of interaction with the real world also means a huge risk of privacy leakage both for individuals and enterprises, which has hindered the wide deployment of metaverse. Then, it is inevitable to apply privacy computing techniques in the framework of metaverse, which is a current research hotspot. In this paper, we conduct comprehensive research on the necessity, taxonomy and challenges when privacy computing meets metaverse. Specifically, we first introduce the underlying technologies and various applications of metaverse, on which we analyze the challenges of data usage in metaverse, especially data privacy. Next, we review and summarize state-of-the-art solutions based on federated learning, differential privacy, homomorphic encryption, and zero-knowledge proofs for different privacy problems in metaverse. Finally, we show the current security and privacy challenges in the development of metaverse and provide open directions for building a well-established privacy-preserving metaverse system. For easy access and reference, we integrate the related publications and their codes into a GitHub repository: \url{https://github.com/6lyc/Awesome-Privacy-Computing-in-Metaverse.git}.
\end{abstract}


%Research highlights
% \begin{highlights}
% \item Anomaly detection
% \item Decoupling anomaly discrimination and representation learning
% \item Contrastive learning and generative learning
% \end{highlights}

% Keywords

\maketitle

% Main text

\section{Introduction}

The rapid development of the Internet has sparked a technological revolution and the rise of metaverse technology. The metaverse, a term that originates from the science fiction \emph{Snow Crash} in 1992, is a virtual world that coexists alongside reality. It leverages Internet technology to establish an immersive and fancy virtual space which is parallel to the physical world, while these two spaces share the consistent time dimension. To be exact, \textcolor{black}{the metaverse combines both the real and the virtualized worlds} \cite{DBLP:journals/access/Al-GhailiKAHOTK22}. It possesses the real-life scenes and also creates new worlds on its own, and each real-life individual is endowed with a virtual avatar, which can interact with users and have the ability to survive on their own in the virtual world. Wikipedia also describes the metaverse as a collective virtual shared space that is created by the convergence of virtually enhanced physical reality and physically persistent virtual space, including the sum of all virtual worlds, augmented reality, and the Internet.

From a macro perspective, the metaverse is not merely a rudimentary virtual world but rather an immensely expansive, intricate and open-ended system. Users are provided with considerable flexibility to conduct any activities, such as gaming, communicating, engaging in social activities, etc., rather than just simulate or mirror the real world. The advantage of the metaverse is brought about by the combination of a series of technologies, including augmented reality (AR) \cite{DBLP:conf/chiplay/Xu22}, virtual reality (VR) \cite{DBLP:journals/jsis/DincelliY22}, digital twin \cite{DBLP:journals/patterns/LvQLYW22}, Internet of Things (IoT) \cite{DBLP:conf/icact/MozumderSAA022}, blockchain technology \cite{DBLP:journals/iotj/FuLYLZL23} and other cutting-edge innovations. These technologies construct a linkage between the online digital world and offline physical world and make the metaverse different from the previous simple virtual world, affording users an immersive experience that transcends spatial and temporal limitations, as though they are physically present. 
As stated by the Meta's CEO, Mark Zuckerberg, the metaverse is a virtual realm where one can fully immerse themselves rather than just looking at it. 
Based on the immersive interaction capabilities, the metaverse is also recognized to be the next successor of the Internet, even though it is a virtual world essentially.

{The applications of the metaverse have extended to all aspects of human society. Especially in the fields of social interaction, tourism, industry and virtual economy, the metaverse has brought great changes to our lives. For example, the metaverse has brought richer sensory experiences for online meetings and games, new ways of traveling during pandemic, new technical support for industry, new operating environments for virtual economy, and so on.}

Generally, the metaverse has six core characteristics as follows. 1) Space-time. The temporal attribute within the metaverse mirrors that of the real world, while the spatial dimension is entirely virtual. 2) Immersiveness. Users are able to perceive a phenomenally realistic and engrossing immersion through extraordinary virtual technology such as augmented reality \cite{DBLP:journals/comsur/WangSZXLLS23}. 3) Real-time. The metaverse boasts high bandwidth and low latency, providing excellent real-time performance. 4) Sustainability. Governed by its own rules and operational logic, the metaverse is self-sustaining and able to maintain itself indefinitely. 5) Interoperability. The metaverse is an interconnected ecosystem that allows for seamless data exchange between different platforms, thereby the digital avatars of users can move to another virtual space freely \cite{DBLP:conf/bigdataconf/ChenWGQ22}. 6) Data sensitivity. Despite the excellent performance, the metaverse will obtain the users' physiological characteristics, such as facial features, height, gender, etc., via the advanced wearable technology and transmit the above information on multiple platforms, which potentially compromises personal privacy and security.

The aim of the metaverse is to break through the existing physical space and then create a new virtual-physical integrated space. 
As a technology integrating physical reality with digital virtual reality, the metaverse has a wide range of impacts on people's lives.
Although it is primarily employed in gameplay \cite{DBLP:conf/isalalife/ChenSLV20}, the significance extends well beyond that. First of all, thanks to its amazing immersive outcome and visual experience, the metaverse is able to bring more entertainment and convenience to human life. People may, for example, indulge themselves in the concert at home, try on clothes online and select the suitable one, travel around and visit the sights of various scenic spots in a computer-generated virtual space. They don't need to go out but can do all kinds of things and experience lots of fun as if they were really there.
Secondly, the metaverse opens up new avenues for education \cite{DBLP:journals/ieicetd/Nagao23, DBLP:conf/bigdataconf/ChenWGQ22}, medical treatment \cite{DBLP:conf/services2/XuJWL22} and humanistic care. It is able to create a realistic teaching scenario and show vivid outcomes that wouldn't be seen normally. Students can learn astronomy in the spacecraft of the virtual world, or simulate chemical experiments and observe the results, which makes education more impressive and reduces costs. In terms of health, wearable devices may help medical professionals examine and monitor patients' physiological data such as body temperature, heart rate and blood pressure. What's more, the metaverse even has the ability to "resurrect" a deceased person using his or her biometric data. 

To date, the metaverse has attracted increasing attention and spawned preliminary applications in both consumer (To C) and business (To B) scenarios. The To C metaverse is initially developed in online games and geared towards individual users. It has gradually stepped out into broader fields including social intercourse, shopping, and media in recent years \cite{DBLP:conf/bigdataconf/ChenWGQ22}, with the aim of bringing about a better immersive experience. The To B metaverse is targeted at enterprises, aiming to realize remote collaboration and improve efficiency and benefits. For example, BMW has built a virtual factory on NVIDIA's \emph{Omniverse} platform to simulate production. Meta constructs an office-social platform named \emph{Horizon Workrooms} to support face-to-face conferences for employees. It is believed that the future potential of the metaverse lies in the To B scenarios, and a growing number of Internet companies are beginning to venture into the To B applications.

However, the booming of metaverse leads to issues about data privacy and security.
Concretely, there exist two potential safety hazards on users' private information. 
On the one hand, both wearable devices and virtual platforms are vulnerable to cyber attacks \cite{DBLP:conf/tpsisa/PietroC21}, posing a risk of data leakage. For instance, the VR eyeglass has become a portal for malware intrusion and privacy leakage, and the scanning devices are able to capture the environment about users' homes. In addition, the virtual avatars contain plenty of identities of real individuals \cite{DBLP:journals/ijsppc/Sebastian23}, such as gender, interest and facial characteristics.
On the other hand, the right of utilization and ownership of the private data do not belong to the individual users, but to the company that manages the metaverse platform. People inadvertently expose much of their private information to the metaverse platforms, but lack control over their data. Companies that own data may sell user information for profit, or utilize it for commercial recommendations, including accurately delivering products that users like and so on.
What's worse, privacy leakage can result in personal financial risk and property loss, as well as being impersonated to engage in criminal activities.
All these give rise to various concerns about the possible abuse of the data, so many people are hesitant to embrace the metaverse.

To protect user privacy and data security, some efforts are made to standardize Internet technologies as well as the metaverse platforms.
In 2021, U.S. bipartisan senators proposed \emph{The Government Ownership and Oversight of Data in Artificial Intelligence Act}, calling for the regulation of data involved in federal artificial intelligence systems, especially facial recognition data. It reflects the cautious attitude of the U.S. Congress towards digital penetration based on data and identity recognition. 
The European governments are highly concerned about the regulation of the metaverse. The \emph{EU Artificial Intelligence Act}, \emph{Digital Services Act}, and \emph{Digital Market Act} released in recent years have introduced a new set of rules for digital services offered by social media, online marketplaces, metaverse and other online platforms. 
The Chinese government also released the \emph{Network Security Law}, \emph{Data Security Law} and \emph{Personal Information Protection Law}, etc., to enhance the data right confirmation and governance of the Internet.
Japan, in 2021, released the \emph{Investigation Report on the Future Possibilities and Issues of the Virtual Space Industry}, which recommends that the government should focus on preventing and solving legal issues occurring in the virtual world, and collaborate with the experts to formulate industry standards and guidelines to regulate the metaverse. 
South Korea takes the lead in forming a metaverse association, aiming to regulate data security and address the ethical and cultural issues associated with the metaverse market. 

Beyond the policy specifications, protecting data security from a technical perspective is also an urgent and necessary way. Privacy computing \cite{PrivacyComputing}, a new trend in the area of Internet security, has developed rapidly and applied to many scenarios that urge for data security. According to the different technical core ideas, the privacy computing technology can be categorized into three branches. The first one is the Trusted Execution Environment (TEE) \cite{DBLP:conf/syscon/Will22}. It seeks to build a reliable and isolated confidential space that is independent of the operating system, hoping to realize the privacy protection on the mobile devices side from the hardware. The second one is Secure Multi-Party Computing (MPC) \cite{DBLP:journals/jaihc/WangLQLGLX15}, which enables confidential calculations to be carried out without revealing personal private data. The last one is Federated Learning (FL), a distributed framework that allows multiple clients to train with their local data and a global model without disclosing the specific data of other participants \cite{DBLP:conf/bigdataconf/TriastcynF19, DBLP:conf/ccs/TruexBASLZZ19}. 
To gain the favor of more users, the metaverse is required to improve its security infrastructure technologically, not just restricting the platform through the policy. As privacy computing has played a paramount important role in data security and privacy protection, it provides a brilliant prospect for the sustainable development of the metaverse.

To promote the development of the metaverse, it is urgent to pay more attention to the potential privacy and security issues behind it. To the best of our knowledge, most of the research \cite{DBLP:journals/jsis/DincelliY22, DBLP:journals/ieicetd/Nagao23, DBLP:journals/iotj/FuLYLZL23} only focuses on the definition, technology and applications of the metaverse. Several papers \cite{DBLP:conf/bigdataconf/ChenWGQ22, DBLP:journals/ijsppc/Sebastian22} discuss the above issues but lack solutions. 
Therefore, in this survey, we aim to emphasize the data security issues of personal information in the metaverse, and summarize several privacy protection techniques applicable to it, which are the most unique highlights compared to other research. The main contributions of this article are as follows:
\begin{itemize}
	\item We make an introduction about the concept, characteristics and related technologies of the metaverse, and conduct a thorough analysis on its development and applications.
	\item We stress the security and privacy issues in the metaverse, and try to fill the gap on the privacy protection techniques. \textbf{We demonstrate how existing privacy computing technology can be applied to protecting the personal data in the metaverse.}
	\item We also outline the challenges that the metaverse encounters and urge companies and researchers to pay attention to these issues. 
\end{itemize}

\section{PRELIMINARY}
\begin{figure}[t]
	\centering
	\includegraphics[width=3.4in]{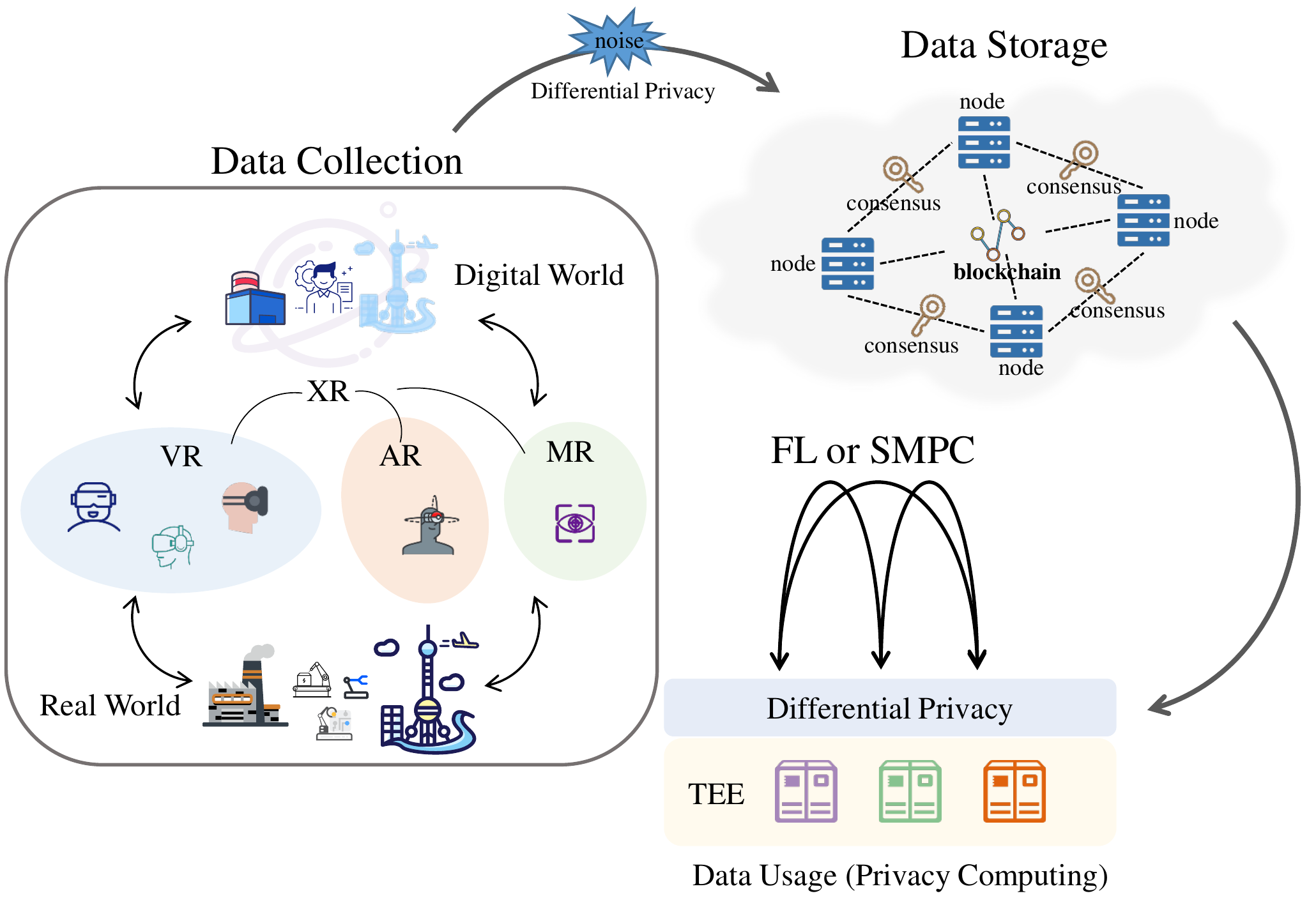}
	\caption{{Key technologies in the metaverse.}}
	\label{fig:preliminary}
\end{figure}

In this section, we will introduce the fundamentals related to Metaverse as shown in Figure \ref{fig:preliminary}, such as extended reality, blockchain, and privacy computing.

\subsection{Extended Reality}
Unlike traditional Internet applications, the metaverse is a digital twin that provides users with ultra-realistic simulations of real-world virtual scenarios with features such as interactive immersion and real-time. Therefore, the construction of the metaverse system requires a terminal carrier that can collect user data with more comprehensive dimensions and realistic feedback. Extended Reality (XR) is a generic term for 3D visual interaction technology which combines real and virtual through computers to provide an interactive virtual environment. XR includes Augmented Reality (AR), Virtual Reality (VR), and Mixed Reality (MR). They become an important technological basis for end carriers in the metaverse.

\subsubsection{Augmented Reality}
The widely accepted definition of AR was proposed by \cite{azuma1997survey}, who argued that AR should have three characteristics: combining the real and the virtual, real-time interaction, and support for three-dimensional registration. The technology simulates and re-exports the human senses of sight, hearing, smell, and touch, and superimposes virtual information on real information to provide users with an experience beyond the real-world sensations. AR systems may include key technologies such as tracking registration, displaying virtual object generation, interaction, and merging virtual and reality. AR has now been extended to urban planning, simulation teaching, surgical treatment, and smart glasses, etc.

\subsubsection{Virtual Reality}
Joe Lanier of the United States proposed VR in the 1980s \cite{Berkman2018}, and it provides users with a multi-information, three-dimensional dynamic, interactive simulation experience by merging computer technology, sensor technology, and so on. Immersion, interactivity, and imagination are the three major characteristics of virtual reality. With the advancement of VR technology, contemporary VR devices may be broadly classified into three types: mobile, all-in-one, and external. The mobile VR device requires the use of cell phones for computing, with the device itself serving just as a display function; the all-in-one VR device has an independent CPU and no external supplementary equipment; and the external VR device consists of a computer, base station, and VR headgear. Because its software features are more sophisticated, VR technology has been used in social, games, movies, and other settings.

\subsubsection{Mixed Reality}
Mixed Reality refers to a novel visualization environment built by mixing technology related to computer vision, graphics processing, display technologies, input systems, and cloud computing. The work of Paul Milgram and Fumio Kishino \cite{milgram1994taxonomy} pioneered MR in 1994. In a nutshell, MR combines the benefits of VR and AR. VR is simply virtual digital graphics, whereas AR is a combination of virtual digital graphics and naked eye reality, and MR is a combination of digital reality and virtual digital graphics. Unlike traditional AR technology, which uses prismatic optics to refract realistic images, MR uses optical perspective technology, video perspective technology, and computer algorithms to produce images in real time, allowing for a more accurate portrayal of AR technology.

\subsection{Blockchain}
Blockchain is one of the primary technologies that contribute to the metaverse's privacy protection, allowing for the secure and trusted storage and distribution of metaverse data. Blockchain is often referred to as a distributed ledger because user data generated in the metaverse can be distributedly stored on blockchain nodes, sensitive data is encrypted, and blocks are linked using cryptographic techniques, ensuring that the data is difficult to tamper with and achieving reliable deposition, data validation, and traceability. Furthermore, a consensus process is utilized to ensure that data or outcomes are consistent among nodes. The smart contract technology in blockchain provides the transparency, openness, and trustworthiness of relevant rule-making and operation, while the incentive system based on smart contracts encourages users to actively participate in maintaining the blockchain.

Blockchain technology contains the following features: fault tolerance, attack resistance, and transparency \cite{gao2018survey}. In terms of fault tolerance, since blockchain systems are essentially distributed architectures, the state of their ledgers is determined by the consensus of most blockchain nodes. Therefore, even the failure of a few nodes will not change the state of the public ledger and will revert to the global public ledger upon re-engaging consensus. In terms of attack resistance, blockchain verifies the validity of added blocks through its underlying consensus mechanism, thus resisting attacks by malicious nodes. Current consensus mechanisms in common use include Proof-of-Work (PoW), Proof-of-Stake (PoS), Practical Byzantine Fault Tolerance (PBFT), etc. In terms of transparency, it is difficult for any node to tamper or delete maliciously because each transaction in the block is executed by all consensus nodes in the consensus mechanism, and all transactions are audited and agreed upon by the majority of nodes.

\subsection{Privacy Computing}
{
Privacy computing is a technology and system for joint computing by two or more participants, who collaborate to perform joint machine learning and analysis without disclosing their respective data. It is a collection of "data available but not visible" technologies.
}
Privacy computing becomes a key privacy-preserving technology underpinning for the metaverse in data utilization, allowing data to be kept local while connecting diverse data silos to speak with one another in order to share data value. Private computing is a synthesis of research and technology from various domains, such as artificial intelligence, cryptography, data science, and so on. 
{
According to the different technical core ideas, privacy computing technologies can be
categorized into four branches: Federated Learning (FL), Differential Privacy (DP), Secure Multi-Party Computing (SMPC), and Trusted Execution Environment (TEE).}

\subsubsection{Federated Learning}
Data is a key driving force behind the quick growth of big data and artificial intelligence. In the real world, the majority of small businesses and organizations struggle with limited data volumes and poor data quality, which makes it challenging to enable the adoption of AI technology. As this is happening, several local regulatory regimes are steadily enhancing data privacy and enacting pertinent laws, like the General Data Protection Regulation (GDPR). In addition, data held by commercial companies or organizations often have great potential value and are reluctant to share data due to issues such as conflict of interest, thus forming data silos. Federated Learning (FL) is a paradigm for cooperative modeling, training, and prediction across data silos under privacy protection. FL \cite{fedavg} was first proposed to address the problem of organizations being unable to share data in order to benefit from it because of problems like competing interests or data security.

Federated Learning can be divided into three main categories based on the data distribution relationship among participants: horizontal federated learning, vertical federated learning, and federated transfer learning \cite{yang2019federated}. For horizontal federal learning, the business or sample features among the participants are similar, and the samples mostly do not overlap, so that the model can learn more knowledge adequately by joint modeling and learning. For vertical federal learning, the samples' ID among participants basically overlap, but the features overlap less, and it is difficult for each party to build an ideal model using only their own business data or a small amount of feature information; therefore, sample alignment techniques and federated learning techniques can be used to assist participants to better use the global model for making decisions. For federated transfer learning, the sample features and samples' ID overlap less among the participants, so the similarity among data, tasks and models can be used to adapt the source domains models to the target domain based on federated learning and transfer learning techniques. In the federated learning process, the medium of transmission and exchange between participants and the server is mostly model parameters or gradients.

{\subsubsection{Differential Privacy}}
{
Differential Privacy (DP) is a privacy safeguard that protects an individual's privacy while allowing relevant information to be derived from data. It provides a mathematical framework and algorithms for limiting the ability of statistical analysis of individuals' data to deduce sensitive information about them.
}

{
The concept of differential privacy was first introduced by Cynthia Dwork et al. in 2006 and has been further developed and generalized in subsequent studies. The basic principle behind differential privacy is to introduce a certain level of noise into the data distribution or data analysis process in order to obfuscate the contribution of individual data and so conceal individuals' identities and sensitive information. By preserving privacy, it tries to avoid data misuse, information leakage, and re-identification threats.
}

{
In the creation and research of differential privacy, several notable publications and methodologies have evolved. To guarantee anonymity, researchers have developed different noise injection approaches, such as the Laplace mechanism \cite{dwork2006calibrating} and the exponential mechanism \cite{chaudhuri2011differentially}, for injecting noise in statistical analysis. Furthermore, differential privacy machine learning methods, such as differential privacy gradient descent \cite{dwork2008differential} and differential privacy generative adversarial networks \cite{abadi2016deep} (GANs), have evolved to allow for effective machine learning while maintaining anonymity. Furthermore, differential privacy's application fields are increasing to encompass healthcare, social networks, financial data, and smart transportation.
}

\subsubsection{Secure Multi-Party Computing}
The theory of secure multi-party computing (SMPC) is a theoretical framework put forth by Yao's ground-breaking work in 1982 to address the challenge of collaborative computation between a group of untrusted participants while protecting information privacy and addressing the lack of a trusted third party. The SMPC can guarantee both the accuracy of the calculation and the privacy of the input while also ensuring that none of the participating parties' input data is disclosed without the involvement of a reliable third party.

Aspects of SMPC that are theoretical include studies on security models, complexity, and viability. The fundamental cryptographic methods used by generic SMPC, which turn target computing activities into arithmetic or boolean circuits, include secret sharing, homomorphic encryption {(HE), zero-knowledge proofs (ZKPs)}, unintended transmission, and obfuscation circuits. While this is going on, researchers may forego some security in order to increase operational efficiency with the SMPC protocol \cite{zhao2019secure}.

\subsubsection{Trusted Execution Environment}
The Trusted Execution Environment (TEE) provides trusted computing with a hardware setting with protective capabilities for securing data and isolating it for processing. The idea is to separate the system's hardware and software resources into two distinct execution contexts, the Common Execution Environment and the Trusted Execution Environment. Both environments have independent internal data pathways and the necessary storage capacity for computing, and they are safely isolated. Even within the TEE, several programs execute independently of one another and cannot be authorized to access or use each other. Apps in the normal execution environment do not have access to the TEE.

TEE supports more arithmetic and complicated computations than SMPC and FL do because it does not place constraints on the algorithmic logic language by computable type. The support for multi-level, highly complicated algorithmic logic implementations and great computing efficiency of TEE are further benefits. TEE is frequently used in conjunction with cryptographic techniques like SMPC to achieve cryptographic protection, which further increases security.

\section{Classification of Metaverse Research and Applications}
\label{sec:CLASSIFICATION OF METAVERSE RESEARCH AND APPLICATIONS}

\begin{figure*}[t]
	\centering
	\includegraphics[width=0.99\textwidth]{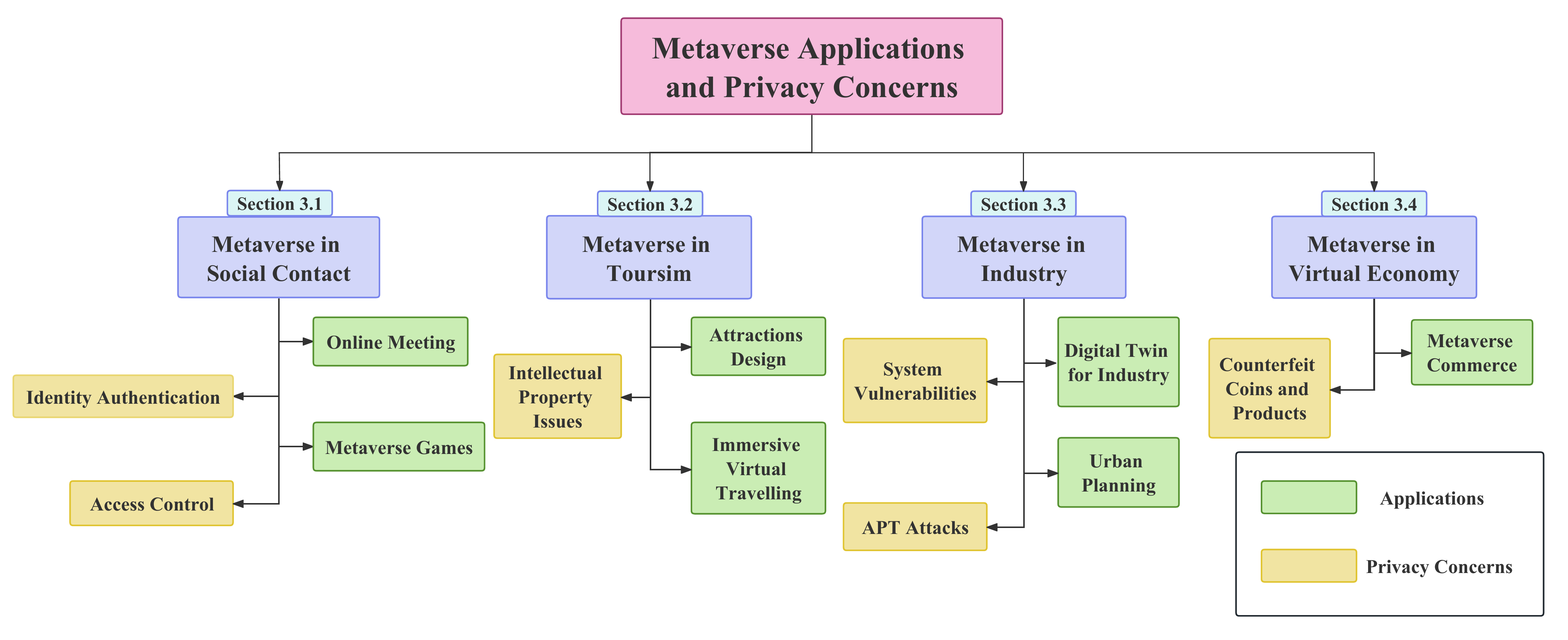}
	\caption{{Metaverse Applications and Privacy Concerns.}}
	\label{fig:metaverse}
\end{figure*}

{The research and applications of the metaverse have spread to many aspects of our lives \cite{li2022metaApp2, li2023metaApp1}.} In areas such as social contact, tourism, industry and virtual economy, the metaverse is beginning to bring significant changes to human society, as shown in Figure \ref{fig:metaverse}. Moreover, due to the wider and deeper applications, the metaverse faces many challenges in data usage.

\subsection{Metaverse in Social Contact}

The Internet has been changing the way people communicate with each other since its appearance. As a new paradigm of the Internet, the metaverse has also brought unprecedented impact to the social field.

The most important technologies for social functions in the metaverse are XR (Extended Reality) and digital avatars. XR devices such as helmet-mounted displays (HMDs) are becoming the main terminals to enter the metaverse \cite{sugimoto2021extended}. Avatars refer to the digital representation of human users in the metaverse \cite{lacey2019cuteness}.

Supported by the above technologies, a number of metaverse social applications have emerged. Many online meetings use augmented reality (AR) technology, which allows users to transform their images into cartoon style. In the virtual world, game players and other participants can modify and edit the appearance of their avatars with almost unlimited options\cite{kolesnichenko2019understanding}, such as the \textit{Fortnite}, a metaverse game. All these features greatly enhance the interactivity of metaverse social.

The social features of the metaverse face many challenges in terms of privacy protection. In the metaverse, identity authentication and access control play a vital role. The identity of users and avatars in the metaverse may be illegally stolen and impersonated. If a user's identity is stolen in the metaverse, his avatar, digital assets and social connections can be leaked and lost. Moreover, the social services of the metaverse generate many new types of personal profiling data, such as biometrics information and daily routine. Malicious attackers may try to illegally gain access to these data, posing a threat to user's privacy.

\subsection{Metaverse in Tourism}

The rise of COVID-19 and the introduction of travel restrictions have changed people's view about travel. As a result, virtual environments are regarded as an alternative for virtual travelling. The metaverse also brings new development opportunities for the tourism industry.

New technologies in metaverse such as VR enable users to travel through computer-generated images and videos that simulate real-world experiences. This has opened up new possibilities for tourist attraction designs. Sweden used VR technology to show construction plans for public comment when planning two roads through cultural heritage areas \cite{heldal2007supporting}. Italy plans a transportation hub using a virtual environment to present it online to the public. Users can explore it as avatars, while interacting with other users and accessing information about the project \cite{caneparo2001shared}.

Nowadays the metaverse is taking virtual travel to a new level, making it more immersive for visitors. Some organizations are already offering virtual 3D versions of real spaces and locations. For example, visitors can currently explore the Louvre Museum in Paris in metaverse. Virtual visitors can view exhibitions, enjoy concerts, and even meet friends at the virtual museum. In addition to virtual reality, the metaverse also uses augmented reality in tourism, which makes the elements in the metaverse not completely virtual, but integrated with the real world around them. A famous example would be the game Pokemon Go\cite{ertel2017qualitative}, where a user sees the titular monsters overlaid in the real world as seen by their phone’s camera.

The metaverse tourism industry faces pitfalls in terms of intellectual property rights as everyone is able to reconstruct tourist attractions in the virtual world and potentially profit from them. For example, some Mediterranean coastal countries have been claiming ownership of monumental images in their countries \cite{addison2007vanishing}. Bangladesh's attempt to build a replica of the Taj Mahal was opposed by India. Chicago banned professional photographers from photographing the city's Millennium Park without permission, claiming that the park is protected by copyright laws\cite{guttentag2010virtual}. Intellectual property issues become more complex in the digital realm. How to define the criteria for intellectual property infringement in the metaverse and safeguard the security and interests of all parties has become an urgent issue.

\subsection{Metaverse in Industry}

The metaverse has also had a profound impact on the industrial field. From automobile manufacturing to air transportation to mining, the metaverse has penetrated into various industrial fields.

The metaverse is essentially a combination of several technologies, and one of the most closely related to industry is the digital twin, which blends digital reality and physical reality\cite{grieves2017digital}. Many ports are already using digital twins to track containers on the dock\cite{klar2023digital}. No matter where they are stacked, we can locate and trace them accurately. Aerospace companies are building engines and airframes in the digital world to simulate how an aircraft will fly before it is actually built. Many new factories exist just as much in the digital world as they do in the physical, allowing operators to visualize operations down to the smallest detail.

Another example of metaverse in industry is urban planning\cite{abouelrous2023digital}. Engineers can roam the streets with their extended reality (XR) glasses and observe intersections through a metaverse lens. They can immediately see the traffic impact of moving a bus stop or adding a traffic light, and then aggregate and upload the proposals to a citywide digital twin that other planners can reference.

While creating new forms of industry, the metaverse also poses a certain threat to the real world. By sniffing software and system vulnerabilities in the metaverse, hackers can use damaged devices in physical industries as entry points to invade national critical infrastructures such as power grid systems and high-speed rail systems through APT (Advanced Persistent Threat) attacks \cite{hu2015dynamic}, posing a great threat to social security.

\subsection{Metaverse in Virtual Economy}

The application of the metaverse in the virtual economy relies on several techniques and concepts.

\textbf{Blockchain.} Blockchain uses proof of work as a consensus mechanism, requiring participants to expend effort on puzzles to ensure data security. The metaverse also requires blockchain and its derivative technologies as a foundation in building a virtual economic system \cite{yang2022fusing}.

\textbf{NFT.} Non-fungible tokens or NFTs are a new set of digital assets based on blockchain technology \cite{wang2021non}. NFT can also be described as a representation of an asset on a blockchain network.

\textbf{Web 3.0.} Web 3.0 (also known as Web 3) is an idea for a new iteration of the World Wide Web which incorporates concepts such as decentralization, blockchain technologies, and token-based economics. Metaverse is considered to be the envolving paradigm of Web 3.0 \cite{grider2021metaverse}. In the metaverse, with the support of physical infrastructure and metaverse engine, users represented by digital avatars can travel between various virtual worlds and experience digital life.

The most representative of the virtual economic system built on the basis of the metaverse is metaverse commerce\cite{lee2021all}. Metaverse commerce is an emerging concept that refers to transactions that occur in the virtual world, including but not limited to user-to-user and business-to-user transactions. Since the transaction process is digital, the transaction system of metaverse commerce can be largely borrowed from the existing e-commerce system. For example, eBay, a representative of C2C e-commerce, can be ported to the metaverse community.

However, metaverse commerce in the virtual economy is not exactly the same as traditional e-commerce. First of all, the items traded are different. The ownership of virtual items should also be effectively protected in the metaverse trading market. For example, Battle Pets \cite{lee2018interaction} and My DeFi Pety \cite{itoh2021towards} allow players to breed and trade their virtual pets. In addition, the focus of metaverse commerce is its interoperability: users can carry digital property in different virtual worlds.

The virtual economy created by the metaverse faces many challenges, and the privacy and security of users' property are of particular concern. Although NFT cannot be occupied by other users of the metaverse community, counterfeit coins may still be generated. For example, after seeing the property of other users on a virtual trading platform, a user with bad behavior may try to create a counterfeit and claim originality to it.

\subsection{Challenges in Data Usage}
The metaverse, as a new paradigm of the Internet, has a wide range of applications in all the above-mentioned aspects. The carrier of information in the metaverse is data, and the use of data faces a variety of challenges.

\subsubsection{Challenges in Privacy and Security}
\paragraph{Security of Interface Devices} The metaverse, as a virtual world parallel to the real world, requires users to access it through various wearable devices\cite{vernaza2012towards}, such as VR glasses, headsets and HMDs (helmet-mounted displays). In order to interact with virtual characters, these devices perform different levels of data collection and behavioral analysis on the user. The information being collected includes biometric features such as facial expressions, body movements, voice, and even brainwave patterns. Once these data are leaked to an attacker, the attacker will be able to achieve user tracking and pose a serious threat to user privacy.

\paragraph{Threats to Identity Authentication and Access Control} Authentication and access control are important in the metaverse. A user's avatar and digital assets may be exposed and lost if their metaverse identity is taken. For example, in 2022, the accounts of 17 users in the Opensea NFT marketplace were hacked due to smart contract flaws and phishing attacks, resulting in \$1.7 million in losses. In addition, because the metaverse requires a high level of interactivity and a large amount of personal information is generated and transmitted in real time, it is complicated to decide exactly what personal information to be shared, with whom, under what condition, and when it is destroyed. Malicious attackers may illegally elevate their data access privileges through buffer overflows, tampering with access control lists, and other methods.

\paragraph{Security of Cloud Storage} Storing privacy-sensitive information in cloud servers and edge devices also poses privacy threats. For example, hackers can query and infer users' private information frequently through differential attacks, or compromise cloud storage through distributed denial-of-service (DDoS) attacks\cite{bertino2017botnets}. For example, the database of \textit{Second Life}, a metaverse game, had been hacked and a large amount of user data, including payment details and passwords, was leaked \cite{wang2022survey}.

\vspace{12pt} % Adjust the value as needed for the desired spacing

{The real-life application of the metaverse has seen many cases of serious repercussions due to unfavorable protection of users' private data. Facebook's owner, Meta, was fined 1.3 billion dollars on May 22, 2023, for mishandling people's data when transferring it between Europe and the United States. The EU regulator said the processing and storage of personal data in the United States contravened Europe’s signature data privacy law, known as the General Data Protection Regulation \cite{Voigt_vondemBussche_2017}. Even as one of the main companies driving the development of the metaverse, Meta has struggled to ensure the privacy of its users, which raises even more concerns about the security of data when the metaverse is widely used.}

\subsubsection{Challenges in Efficiency}

Metaverse has strong social properties and its applications are usually multi-user, such as multi-player games and remote collaboration. How to achieve secure and efficient content sharing in XR environment in metaverse becomes a challenge in data usage. In addition, the sharing and processing of user-generated content (UGC) in the metaverse is also important. How to reduce the communication burden without affecting content validation is also a challenge for the metaverse.

\subsubsection{Challenges in Data Heterogeneity}

The heterogeneity of the data is caused by the nature of the multiple parties involved in the metaverse. This poses a challenge for data usage. The heterogeneity of the metaverse includes heterogeneous virtual spaces due to different implementations, heterogeneous physical devices due to different interfaces, heterogeneous data types, heterogeneous communication methods, and so on. This also makes the interoperability of the metaverse difficult to implement.

\begin{figure*}[t]
	\centering
	\includegraphics[width=1\textwidth]{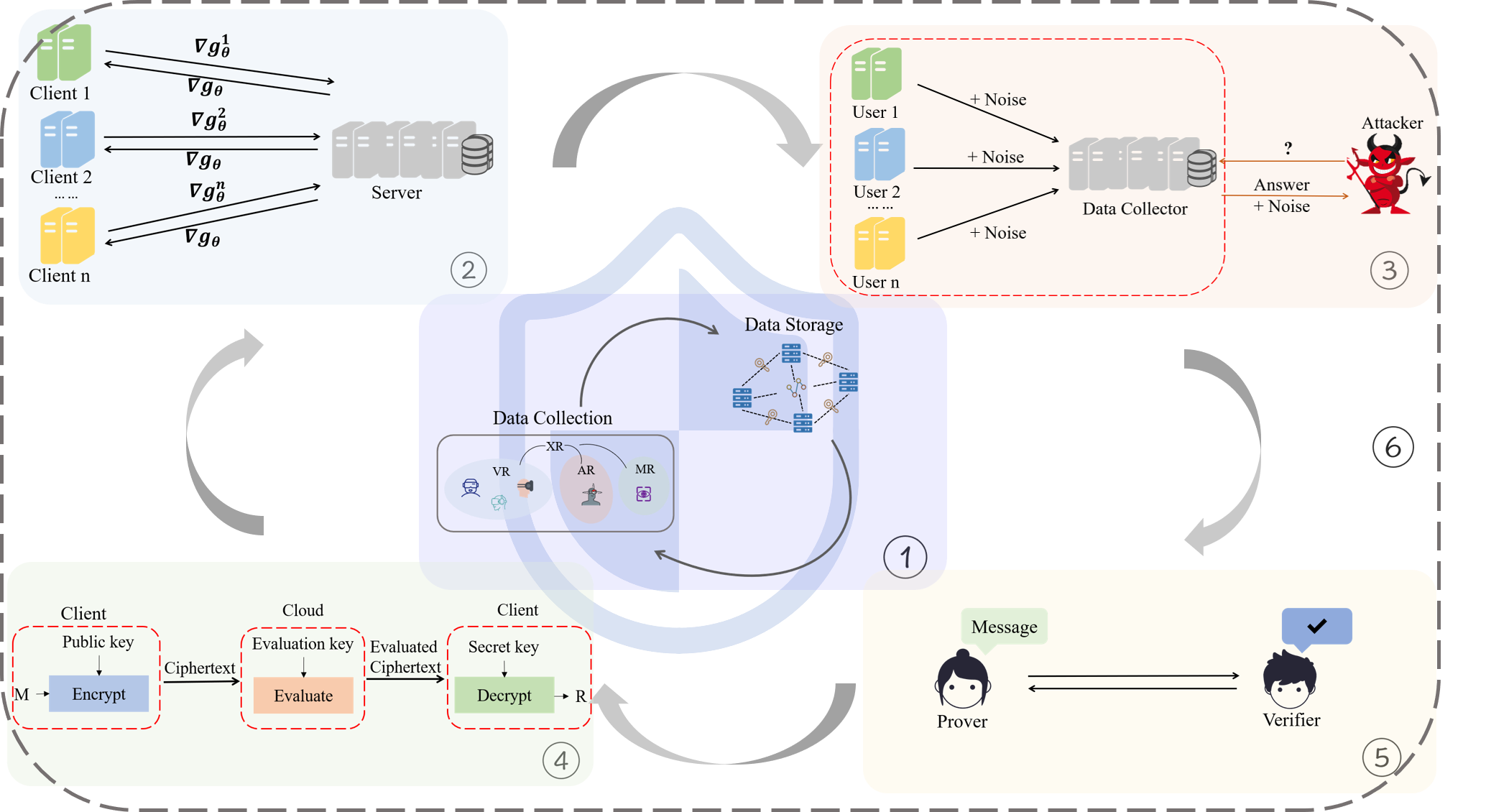}
	\caption{{The overview of privacy computing in the metaverse.} \ding{172} is the schematic diagram of \textbf{Metaverse}. {It relies on various terminals, such as VR and AR, to collect user personal information and environmental data, and then store them on the server. The server analyzes and processes this information, and then transmits feedback to the terminal device.} \ding{173} is the schematic diagram of \textbf{Federated Learning} ($\nabla g_\theta^i$ represents the model gradient of each client). \ding{174} is the schematic diagram of \textbf{Differential Privacy}. \ding{175} is the schematic diagram of \textbf{Homomorphic Encryption}. The letters M and R indicate \textit{Message} {(i.e. key and sensitive data)} and \textit{Result} respectively. \ding{176} is the schematic diagram of \textbf{Zero-Knowledge Proofs}. \ding{177} is the chematic diagram of \textbf{Trusted Execution Environment}. These five privacy computing technologies constitute a comprehensive privacy-preserving system for the metaverse. {The interconnectedness of these components lies in their collective contribution to a privacy-preserving system. Data is protected through decentralized training (FL), additional noise (DP), encryption (HE), proofs without revelation (ZKPs), and secure execution (TEE), ensuring that users can interact within the metaverse while minimizing the risk of privacy breaches.}}
	\label{fig:privacy_computation_in_metaverse}
\end{figure*}

\section{WHEN METAVERSE MEETS PRIVACY COMPUTING}

\begin{table*}[t]
	\centering
	\caption{Summary of Different Privacy Computing Technologies In The Metaverse: Method and Benefits.}
        \resizebox{2.0\columnwidth}{!}{%
	\begin{tabular}{|c|c|l|}
		\hline
		\textbf{Privacy Computing} & \textbf{Ref.} &
		\makecell[c]{\textbullet \ \textbf{Proposed Method} \\
			\textopenbullet \ \textbf{Specific Benefits}}
		\\ 
		\hline
		\multirow{6}{*}{Federated Learning} & \cite{zhou2022resource_allocation} &  
		\makecell[l]{
			\textbullet \ Combinatorial optimization problem construction and resource allocation algorithm design \\
			\textopenbullet \ Tradeoff between energy, execution delay, and model accuracy}
		\\ \cline{2-3} 
		
		& \cite{zeng2022hfedms} &
		\makecell[l]{\textbullet \ Dynamic sequential-to-parallel training strategy for heterogeneous streaming data \\
			\textopenbullet \ Robustness to heterogeneous data in the industrial Metaverse}
		\\ \cline{2-3}
		
		& \cite{jiang2021digital_twin} & 
		\makecell[l]{\textbullet \ Incentive mechanism based on Iterative double auction \\
			\textopenbullet \ More aggressive local updates and better quality global model}
		\\ \cline{2-3} 
		\hline
		
		\multirow{6}{*}{Differential Privacy} & \cite{wei2020ldp_res} & 
		\makecell[l]{\textbullet \ Graph-based LDP algorithm and dynamic graph-based CSI algorithm for topic recommendation \\
			\textopenbullet \ Strong privacy protection for both local and online social content}
		\\ \cline{2-3} 
		
		& \cite{zhang2021privacy_mobile} & 
		\makecell[l]{\textbullet \ Data aggregation and data auction with differential privacy for mobile crowdsensing \\
			\textopenbullet \ Tradeoff between data privacy and model accuracy}
		\\ \cline{2-3}
		
		& \cite{nair2022dp_vr} & 
		\makecell[l]{\textbullet \ Dynamic differential privacy mechanism for user requirements and open source Unity plugin \\
			\textopenbullet \ More flexible privacy levels and high-fidelity VR effects}
		\\ \cline{2-3} 
		\hline
		
		\multirow{4}{*}{Homomorphic Encryption} & \cite{chen2018armor} &
		\makecell[l]{\textbullet \ Dynamic spectrum allocation with privacy protection \\
			\textopenbullet \ Comprehensive protection for sensitive user information and more efficient spectrum allocation}
		\\ \cline{2-3} 
		
		& \cite{ma2022fl+he} & 
		\makecell[l]{\textbullet \ Multi-key Homomorphic Encryption Protocol design for federated learning \\
			\textopenbullet \ Better defense performance against malicious clients and servers}
		\\ \cline{2-3} 
		\hline
		
		\multirow{4}{*}{Zero-Knowledge Proofs} & \cite{ghirmai2023ssi} &
		\makecell[l]{ \textbullet \ Blockchain framework construction based on self-sovereign identity \\
			\textopenbullet \ Better solution of decentralization, reliability and interoperability problems in the metaverse}
		\\ \cline{2-3} 
		
		& \cite{babel2023zkps} & 
		\makecell[l]{ \textbullet \ General-purpose zk-SNARK protocol for digital wallets \\
			\textopenbullet \ Integration of scalable revocation, certificate linking and secure element}
		\\ \cline{2-3} 
		\hline

            \multirow{4}{*}{{Trusted Execution Environment}} & \cite{xu2023trustless} &
		\makecell[l]{ \textbullet \ {A blockchain-enabled metaverse based on trust evaluation} \\
			\textopenbullet \ {Efficient resource integration and allocation, flexible and trusted computing environment} }
		\\ \cline{2-3} 
		
		& \cite{liu2022TEEoffchain} & 
		\makecell[l]{ \textbullet \ {TEE-based on-chain and off-chain trusted blockchain}\\
			\textopenbullet \ {Low-cost, high-security execution environments and consistency protocols that facilitate trust extension
}}
		\\ \cline{2-3} 
		\hline
		
	\end{tabular}%
        }
	\label{privacy-computation in metaverse}
\end{table*}

It is clear from the introduction in Section \ref{sec:CLASSIFICATION OF METAVERSE RESEARCH AND APPLICATIONS} that the metaverse is growing rapidly and is becoming more and more connected to the real world. Meanwhile, this means that there may be serious risks of privacy leakage in terms of data usage in the metaverse. Therefore, the study of privacy computing in the metaverse is necessary and also gradually becoming popular. Privacy and data security are the foundation of metaverse construction, and only by solving the privacy protection problem can participants create and interact with more peace of mind in the metaverse, thus promoting the sustainable development of the metaverse. {In this paper, we conduct an exhaustive survey of state-of-the-art privacy computing methods that are widely used in the metaverse. In particular, we focused on privacy computing technologies closely related to the underlying foundation of the metaverse (such as XR, blockchain, etc.), as well as the practical applications of these technologies in the metaverse (such as topic recommendations, digital wallet, AI-generated content, etc.). We also studied privacy computing technologies related to challenges (such as communication delays, data heterogeneity, etc.) faced by the metaverse industry.}

In the following, we will introduce the work of privacy computing in the metaverse as four aspects, summarized in Table \ref{privacy-computation in metaverse} and as shown in Figure \ref{fig:privacy_computation_in_metaverse}. 

\subsection{Federated Learning in Metaverse}
Federated Learning (FL) is a distributed approach of machine learning. It allows collaborative model training of all parties by uploading gradients without exposing private data \cite{mcmahan2017FL}. It is a good paradigm for solving the privacy problem in the metaverse. {For example, in Federated Learning (FL) applied to Augmented Reality (AR), Virtual Reality (VR), or Extended Reality (XR), the training typically involves model updates based on user interactions and experiences within these immersive environments. The data used for training in FL is decentralized and remains on the user’s device. The model is trained collaboratively across multiple devices without the need to centralize raw user data. In an AR application, each user’s device may collect data on their interactions with augmented content, preferences, or navigation patterns. In FL, instead of sending this raw data to a centralized server, the model is updated locally on each user’s device. Only the model updates (usually in the form of gradients) are sent to a central server, where they are aggregated to improve the global model. This process helps to preserve user privacy by keeping sensitive data on the user’s device.} However, there are still problems of resource allocation, data heterogeneity, and lack of user motivation in the scenarios applied to the metaverse.

First of all, the metaverse requires a large amount of resources in several aspects, including computing resources, storage resources, bandwidth resources, etc., due to its large number of users and the wide geographical area involved. Especially for various edge devices, these resources are very poor. For example, AR applications rely on efficient perception and computation of the real world, as well as real-time rendering on AR display devices, which requires a substantial amount of computational resources to function properly. Insufficient computational resources can cause issues such as degraded device performance, increased latency, and even failure to operate. {The authors in \cite{chen2020fl+ar} first propose a paradigm framework that combines federated learning with mobile edge computing for AR applications.} It allows for the acquisition of globally optimal machine learning models with fewer training rounds and less communication consumption than centralized training, resulting in significant savings of computational and bandwidth resources. {Furthermore, in order to balance energy consumption, execution latency, and model accuracy of AR applications in different scenarios within the metaverse, the authors in \cite{zhou2022resource_allocation} construct a non-convex optimization problem and design a resource allocation algorithm to determine the bandwidth allocation, transmission power, CPU frequency, and video frame resolution for each client in FL framework.} In addition, the Industrial Metaverse aims to integrate the physical and digital worlds to improve the efficiency and safety of industrial production. To address the latency problem caused by the low transmission rate of cellular-based Low Power Wide Area Networks (LPWAN), it proposes HFEDMS \cite{zeng2022hfedms} that combines two model parameter synchronization modes to assign more communication rounds to the more important and lightweight classifier parameters, which reduces the communication cost while sub-assuring the overall performance. 

Secondly, in the system of metaverse, the data sources are multiple. They can be personal data provided by different users, such as behavioral trajectories, social relationships, etc., or various environmental data collected by different sensors in the virtual as well as real world. These heterogeneous and unbalanced data can greatly disrupt the training effect of federated learning \cite{zhao2018noniid}. The authors in \cite{zhang2021fedsens} addresses the heterogeneity and class imbalance of health data collected by each end device in an abnormal health detection (AHD) system. It proposes a new federated learning framework called FedSens, which combines reinforcement learning strategies to guide the selection of local clients for high-quality local updates, thus improving the accuracy of the global model. Similarly, data in the industrial metaverse system is often highly dynamic and heterogeneous. \cite{zeng2022hfedms} introduces a dynamic training mode called Sequential-to-Parallel (STP) that can adapt to the continuously changing streaming data and reduce the effect of data heterogeneity.

In the metaverse system, the overall motivation of individual participants is difficult to ensure due to their different goals, resources, and technology levels. Therefore, some incentives are needed to encourage them to actively participate in the construction and development of the metaverse. These incentives can offer various rewards for participants to encourage them to create and share content, engage in social interactions, and provide services in the metaverse. For example, {the authors in \cite{li2020blockchain_committee} propose a decentralized federated learning framework based on blockchain with an incentive algorithm named profit sharing by contribution, which combines two parts: permission fee and profit sharing.} Its convergence guarantee is proved in \cite{che2022blockchain_committee_convergence}. Moreover, the authors in \cite{kang2022blockchain_aoi} design an age-based contract model to incentivize data awareness among different Industrial Internet of Things (IIoT) nodes, thus improving the quality of service in the industrial metaverse framework. {In the digital twin scenario, the authors in \cite{jiang2021digital_twin} present a new blockchain-empowered digital twin edge network framework and a federated learning approach based on Iterative double auction.} Through the Iterative double auction incentive mechanism, the participants can decide the bids based on their data contributions and values to obtain higher revenue, and also motivate more participants to actively perform local model updates to get better quality global models.

\subsection{Differential Privacy in Metaverse}
Differential Privacy (DP) is a common tool for sharing data in distributed machine learning, which protects sensitive personal information by adding random noise to the data locally or in the server \cite{abadi2016DP}. In the metaverse, users often need to upload large amounts of personal data to the server, which requires differential privacy mechanisms to protect the privacy of the individuals involved. The authors in \cite{bi2020DP_edge_Computing} apply local differential privacy to edge computing. They protect the privacy-sensitive location data of each user by randomly disturbing the Voronoi grid where the edge nodes are located. To provide a privacy-preserving trending topic recommendation service in a metaverse, the authors in \cite{wei2020ldp_res} propose a local DP-based algorithm that combines a graph-based LDP (GLDP) algorithm with a dynamic graph-based CSI (DGCSI) algorithm to achieve local privacy recommendations while protecting the privacy of online social content. Moreover, Mobile Crowdsensing (MCS) refers to large-scale data collection and processing based on mobile devices that can provide real-world data to support scenes in the metaverse. The authors in \cite{zhang2021privacy_mobile} propose an auction mechanism for mobile crowd-sensing with differential privacy data aggregation, namely DPDA and EDPDA. In addition, {to address privacy issues in VR, the authors in \cite{nair2022dp_vr} present the first algorithm to implement an incognito mode for VR. It intelligently adds differential privacy noise of different sizes when and where it is most needed and achieves a balance between privacy protection and model effects.} They have also implemented its algorithm as a generic Unity plugin and experimented with it on several popular VR applications, all with satisfactory privacy-preserving results. {In particular, DP can be applied independently in the metaverse, and it can also be integrated with FL to achieve a more advanced privacy-preserving distributed machine learning framework. Therefore, DP and FL coexist as privacy computing technologies within the metaverse.}

However, the application of differential privacy would introduce additional noise that affects the usability of the model, also in the metaverse framework. There has been relatively little research focusing on recovering the performance of the model in a metaverse scenario after using differential privacy techniques, and this is a point of interest for future research.

{\subsection{Secure Multi-Party Computing in Metaverse}
In this section, we choose to delve into two cryptographic techniques closely related to the metaverse called homomorphic encryption (section \ref{sec: HE}) and zero-knowledge proofs (section \ref{sec: ZKPs}), which are included in the context of secure multi-party computation.}

\subsubsection{Homomorphic Encryption in Metaverse} \label{sec: HE}
Homomorphic Encryption (HE) is a common technique for encrypting data in cloud computing, making the result of an operation on a ciphertext the same as an encrypted plaintext operation, which maintains the computable and invisible nature of the data \cite{gentry2009HE}. Compared to differential privacy, homomorphic encryption does not affect the utility of the model and is naturally applied to protect private data in the metaverse. Among them, \cite{saxena2023HE_cloud} designs a secure cryptographic gateway based on homomorphic encryption for cloud computing and analyses the security, complexity, and robustness of the proposed model in this paper. Dynamic Spectrum Allocation (DSA) is a radio communication technology that solves the problem of insufficient network resources in the metaverse by dynamically allocating available spectrum resources to each user. The authors in \cite{chen2018armor} propose the first combined auction framework for heterogeneous spectrum with privacy protection, known as ARMOR, which fully protects users' personal location information through algorithms such as homomorphic encryption. In addition, to protect the privacy of user information in the IoT, the authors in \cite{ma2022fl+he} design a multi-key homomorphic encryption protocol called xMK-CKKS, which yields a more privacy-preserving federation learning framework that prevents data leakage due to collusion between most clients and servers. However, for large-scale computing tasks in the metaverse, homomorphic encryption has the drawbacks of low computational efficiency and high computational cost. In the future, we can continue to explore the application of homomorphic encryption in the metaverse from the aspects of hardware acceleration and optimized algorithm design, making it more suitable for the needs of large-scale operations.

\subsubsection{Zero-Knowledge Proofs in Metaverse} \label{sec: ZKPs}
Zero-Knowledge Proofs (ZKPs) is a cryptographic technique that allows a person to prove that he possesses certain information without revealing any details of that information \cite{goldwasser2019ZKPs}. In the metaverse, zero-knowledge proofs techniques are widely used for user identity or data verification. For example, the authors in \cite{zero_ssi} propose that privacy-enhancing technologies such as zero-knowledge proofs can be used for the reading and verification of digital credentials. \cite{ghirmai2023ssi} talks about two criteria related to the metaverse and builds a blockchain framework using Self-Sovereign Identity (SSI) to enable privacy-preserving meta-universe interoperability. Further, to address the lack of scalable revocation, certificate linking, and integration with secure elements in SSI, \cite{babel2023zkps} implements general-purpose ZKPs (zk-SNARKs) that can also provide scalable and flexible privacy for SSI. In addition to identity verification, zero-knowledge proofs are widely used in other parts of the metaverse. For example, {the authors in \cite{guan2020blockmaze} propose BlockMaze, the first privacy-guaranteed zk-SNARKs-based account model blockchain, which protects account balances, transaction amounts, and information about interactions between traders. The authors in \cite{lin2023Secure_Semantic} combine blockchain and zero-knowledge proofs to defend against malicious semantic attacks for AI-generated content (AIGC) in the metaverse.} Firstly, they propose a blockchain-based semantic communication framework as a way to reduce the redundancy of transmitted information for the AIGC service scenario. Then, zero-knowledge proofs technique is used to verify the specific transformation of semantic data by honest edge devices, effectively preventing the attacks of malicious devices. On the other hand, to improve the efficiency of the authentication algorithm, the authors in \cite{yang2022secure_authentication} construct a chameleon collision signature method that eliminates zero-knowledge proofs while having strong privacy guarantees.

{\subsection{Trusted Execution Environment in Metaverse}
Trusted Execution Environment (TEE) is a computing environment that provides a certain level of security to ensure that programs and data running in it are protected. This environment is usually implemented at the hardware level or through a combination of hardware and software, which is characterized by high performance and real-time response. In the metaverse, TEE is usually combined with the blockchain to jointly create a strictly trusted user interaction environment. For example, the author in \cite{xu2023trustless} proposed using a hypergraph to model the metaverse, and then evaluate the trustworthiness of each group of users based on graph analysis, thereby establishing a trusted environment using blockchain as the underlying technology based on a trust evaluation system. Furthermore, to ensure consistent security on and off the chain, \cite{liu2022TEEoffchain} proposed a trusted environment monitoring system and consistency protocol that supports the TEE to extend the trust of the blockchain from the on-chain to the off-chain.}

\subsection{Lessons Learned}
\begin{enumerate}
	\item \textit{Federated learning cannot meet the privacy needs of the metaverse framework}: By transferring gradients rather than data, federated learning provides a nice privacy-preserving paradigm for the usage of data in distributed machine learning. Existing work has conducted extensive research on issues such as resource constraints, data heterogeneity, and insufficient participation of all parties in the metaverse, which promotes the application of federated learning in the metaverse framework. However, privacy issues in the metaverse still cannot be fully addressed using federated learning techniques alone. For example, a gradient leakage algorithm is proposed in \cite{zhu2019DLG}, which can precisely recover the original private data from the shared gradients. This means that we need to combine multiple privacy computing techniques to build a complete privacy protection system in the metaverse, as shown in Figure \ref{fig:privacy_computation_in_metaverse}.
	\item \textit{Differential privacy would affect model performance while protecting data}: In general, differential privacy techniques protect the privacy-sensitive information in the metaverse data by adding random noise to it. However, the added noise disrupts the original data distribution, thus degrading the performance of each module in the metaverse. To reduce the impact of DP noise on the federated learning framework, the authors in \cite{shen2022performance_DP} optimize the robustness of the model by adding a regular term in the local update. Furthermore, the authors in \cite{li2024FedCEO} propose a flexible differentially private federated learning algorithm based on tensor low-rank optimization, establishing a utility-privacy trade-off framework with theoretical guarantees. In the metaverse, we can also compensate for the negative impact of differential privacy techniques in related ways, which is a frontier research direction.
	\item \textit{The computational complexity issue of homomorphic encryption}: It is clear from our research that homomorphic cryptography strictly protects the privacy data in the computation of metaverse, which is guaranteed by the sound mathematical theory. However, the huge computational effort of the encryption as well as decryption processes greatly slows down the diffusion of homomorphic encryption techniques in the metaverse framework. How to improve the existing homomorphic encryption scheme to make it better applicable in metaverse is a question worth thinking about.
	\item \textit{The usability issues of zero-knowledge proofs}: In the metaverse, zero-knowledge proofs play an important role in the verification of identity rights, transactions, and assets of virtual users. It can verify the authenticity of users' information without accessing their particular data. However, the application of zero-knowledge proofs also consumes a large amount of computational resources. Moreover, zero-knowledge proofs require multiple verifications and authorizations from users, which may affect the user experience in the metaverse.
        {\item \textit{The scalability issues of Trusted Execution Environments}: Firstly, the implementation of TEE usually requires hardware support, which may make it difficult to deploy on some devices, especially those with limited resources. Secondly, although there are some common TEE implementations, such as Intel SGX \cite{costan2016intel_SGX} and ARM TrustZone \cite{pinto2019ARM_TrustZone}, there is a lack of unified standards within the industry, which may lead to interoperability and portability issues.}
\end{enumerate}

Both metaverse and privacy computing are hot topics of current research. The combination of metaverse and privacy computing technologies not only solves various privacy leakage risks in the metaverse but also gives new life to the field of privacy computing.

\section{THE FUTURE DIRECTION}
The metaverse is integrated with many high technologies such as high-speed networks, the Internet of Things, AR, VR, cloud computing, edge computing, blockchain, artificial intelligence and others. With the support of these technologies, the metaverse may potentially provide us with a space that connects the physical and digital worlds. In the development of the metaverse, security and privacy issues are so important that cannot be ignored. If adequate security and privacy protection cannot be provided, the use of the metaverse could even result in significant losses for users in the actual world. Therefore, there are still security and privacy-related issues that need to be resolved in the future in order to promote the further development of the metaverse. 
{
We investigate the bottlenecks of privacy computing in the metaverse and summarize
future directions into three categories: 
\begin{itemize}
    \item Identification, privacy, and security without trusted third-party supervision (Section 5.1 - section 5.4).
    \item The tradeoff between protection and user experience (Section 5.2 - section 5.5).
    \item The corresponding impact on the real world (Section 5.6 - section 5.7).
\end{itemize}
The possible research directions are listed as follows:}

\subsection{Identification and Control of AIGC}
% {In such a decentralized scenario, Without trusted third-party supervision, the identification and control of AIGC is quite challenging.}
The relatively simple acquisition of AIGC is due to the generative models' quick development. But still, it is quite challenging to recognize and manage AIGC. Malicious actors may utilize AIGC to falsify verification or to spread rumors\cite{AIGC}. The difficulty of standardizing the deployment of AIGC in the Metaverse has increased due to the heterogeneity of AIGC identification and control capabilities between users. It is crucial to study how to recognize and manage AIGC in the Metaverse.

\subsection{Privacy and Security Protection Technology}
The protection of security and privacy is always an important issue in the application of the metaverse. If effective security and privacy protection cannot be provided, it may lead to the leakage of users' personal information, resulting in huge losses. For example, real-time location leakage when users use AR navigation, and personal identity information leakage when using VR applications for entertainment.

A range of privacy and security protection methods have been extensively used in the development of the mobile internet, such as the homomorphic encryption \cite{zhang2020batchcrypt,yi2014homomorphic,paillier2005paillier,fang2021privacy}, the differential privacy \cite{abadi2016deep,dwork2006differential}, and the data anonymization \cite{ghinita2007fast,ghinita2009framework}. However, when employed in a metaverse scenario, these techniques can fall short of the requirements {that they are not applicable for decentralized framework or they might bring bad user experience}. 
The data anonymization techniques may disrupt the scenarios in the metaverse and significantly degrade the user experience.
Although differential privacy can protect privacy to some extent, it also reduces the availability of data. Moreover, there are numerous devices in the metaverse, and finding a suitable differential noise intensity for each device is a highly challenging task.
The homomorphic encryption provides strong security and privacy for data analysis in cloud computing environments. However, the homomorphic encryption also incurs a huge computational overhead that may limit its scalability and performance in the metaverse. 
{The experiments conducted by NVIDIA Clara demonstrate that federated learning with homomorphic encryption results in approximately a 15x increase in communication overhead and takes an extra 20\% training time \cite{nvidia}.}
Therefore, developing a privacy and security protection technology with high confidentiality and acceptable computational overhead is a worthwhile research direction for the development of the metaverse.

\subsection{Endogenous Security Mechanisms}
To protect against malicious attacks, it is necessary to fix bugs, use the latest firewall and regularly update applications.
Applications and firewalls are extremely likely to be broken by new attack techniques, causing large losses, if security upgrades are not applied in a timely manner.
The security upgrades to firewalls and applications are typically implemented through frequent patches in the mobile Internet age \cite{altekar2005opus,cavusoglu2008security}. Such a way of updating calls for a service center and requires a lot of communication overhead.
The metaverse has a decentralized architecture that is incompatible with the common method of frequent patch updates. This method depends on a service center to store and distribute the secure service packs. Therefore, it is crucial to develop a security update mechanism that fits the metaverse situation and that supports decentralized self-security updates.

\subsection{Detection of Malware and Malicious Devices}
The development of the metaverse is inseparable from the support of a large number of software programs and hardware devices. Therefore, it is necessary to develop effective measures to cope with the threat of malware and malicious devices.
User interactions are far more private in the metaverse than they are in traditional sceneries such as phone calls and online videos.
In the metaverse, people can easily steal private information with a high level of privacy if they employ malware, and it is simple to act as a relay node for malware to attack other users \cite{vondravcek2023rise}.
Similarly, people can also steal the user's private information by some malicious devices (such as malicious sensors and wireless transceiver devices), and even launch serious attack methods such as wormholes \cite{kuo2023metaverse}, making it difficult for users to detect when they are attacked, and eventually cause incalculable losses.

\subsection{Metaverse Security Architecture}
The operation of the metaverse relies on the integrated collaboration of multiple technologies. Developing an efficient metaverse architecture with sufficient privacy and security protection is a problem worth researching.
Blockchain is considered to be a potential metaverse platform that enables a decentralized tamper-proof and secure computing environment \cite{ersoy2023blockchain,xu2023trustless,lim2022realizing}. However, the blockchain still suffers from some shortcomings. The blockchain-based metaverse systems require an enormous amount of computing resources, which could result in a decrease in user experience.
Besides, many services incorporated into the metaverse have varying QoS/QoE requirements. Thus, it is also a challenging problem how to allocate flexible computing resources for various services. Moreover, blockchain also faces threats of Distributed Denial of Service (DDoS) attacks within peer-to-peer networks, such as the common eclipse attack. Recently, \cite{erfan2023blockchainCD} proposes an intrusion detection system based on a community detection algorithm \cite{li2023CD1, li2023CD2} that aims to identify and thwart attempts to launch eclipse attacks in blockchain-enabled Metaverse systems.

\subsection{Authentication and Supervision}
The information in the actual world is tightly bound to that in the digital world owing to the metaverse \cite{zhang2022llakep,xu2022metaverse,yang2022secure,chakkaravarthy2023metasecure}.
If someone, unfortunately, suffers from information theft in the metaverse, the person is highly possible to be affected directly in the real world.
In addition, it's essential to maintain surveillance on the user's behavior, ensuring that the user behaves as much as possible in line with the relevant authority, and making ensure that their illegal behaviors are traceable and private.
Therefore, it is extremely important to design a decentralized authentication method to effectively verify and govern the digital lives in the metaverse.

\subsection{Privacy Protection Policies}
Although the metaverse further breaks the physical isolation, it still needs to comply with real-world laws and regulations, even though the laws and regulations on user privacy protection in various countries and regions around the world may differ substantially.
In order to achieve further development of the metaverse, it is necessary for all countries and regions in the world to work together to formulate a passable privacy protection policy. 
Also, the service providers are supposed to develop the metaverse services according to the users of each country and region, complying with the local laws and regulations.

\section{Conclusion}
This paper investigates how privacy computing relates to the metaverse. 
We first review the development and privacy protection policies of the metaverse, and discuss why privacy computing is feasible and necessary for the metaverse.
Next, we provide an overview of the technologies that enable the metaverse, such as VR, AR, XR, blockchain, and privacy computing. 
Then, we review the current research on the metaverse in different fields, such as social interaction, tourism, industry, virtual economy, etc., and highlight the data challenges, such as Security of Interface Devices, Threats to Identity Authentication and Access Control, and Data Heterogeneity that emerge in these contexts.
Furthermore, we categorize and evaluate the existing research on privacy computing in the metaverse including federated learning, differential privacy, homomorphic encryption, and zero-knowledge proofs.
Finally, we summarize the main challenges and the future directions of privacy computing in the metaverse. 
We hope this paper will contribute to a better understanding of privacy computing and inspire more outbreaking research in the metaverse.

\printcredits

\textbf{Declaration of interests}
 
The authors declare that they have no known competing financial interests or personal relationships that could have appeared to influence the work reported in this paper.

\textbf{Data availability}

No data was used for the research described in the article.
\bibliographystyle{cas-model2-names}
%\bibliographystyle{elsarticle-num}

\bibliography{paper}

\end{document}